# Decoding Neural Responses in Mouse Visual Cortex through a Deep Neural Network

Asim Iqbal
Brain Research Institute, UZH
Neuroscience Center Zurich
(ZNZ), UZH/ETH Zurich
Zurich, Switzerland
iqbal@hifo.uzh.ch

Phil Dong
Department of Neuroscience
Icahn School of Medicine at
Mount Sinai
New York, USA
zhe.dong@icahn.mssm.edu

Christopher M Kim
National Institute of Diabetes and
Digestive and Kidney Diseases
National Institutes of Health
Bethesda, Maryland, USA
chrismkkim@gmail.com

Heeun Jang
Buck Institute
for Research
on Aging
Novato, CA, USA
hjang@buckinstitute.org

*Abstract*— Finding a code to unravel the population of neural responses that leads to a distinct animal behavior has been a long-standing question in the field of neuroscience. With the recent advances in machine learning, it is shown that the hierarchically Deep Neural Networks (DNNs) perform optimally in decoding unique features out of complex datasets. In this study, we utilize the power of a DNN to explore the computational principles in the mammalian brain by exploiting the Neuropixel data from Allen Brain Institute. We decode the neural responses from mouse visual cortex to predict the presented stimuli to the animal for natural (bear, trees, cheetah, etc.) and artificial (drifted gratings, orientated bars, etc.) classes. Our results indicate that neurons in mouse visual cortex encode the features of natural and artificial objects in a distinct manner, and such neural code is consistent across animals. We investigate this by applying transfer learning to train a DNN on the neural responses of a single animal and test its generalized performance across multiple animals. Within a single animal, DNN is able to decode the neural responses with as much as 100% classification accuracy. Across animals, this accuracy is reduced to 91%. This study demonstrates the potential of utilizing the DNN models as a computational framework to understand the neural coding principles in the mammalian brain.

## I. INTRODUCTION

In the class of mammals, mice are nocturnal animals with a well-organized somatosensory (barrel) cortex that encodes sensory information of 3D environment through their whiskers [1]. These barrels are anatomically and functionally organized in a somato-topic manner [2] to encode the sensory perception from a variety of body parts. These whiskers act as the 'eyes' of mouse especially in the dark [3] as these animals have evolved over time in a similar set of environments. In contrast, primates' visual cortex is hierarchically organized in a visuo-topic fashion, and hence it is widely studied as a classical model to understand the information processing of visual stimuli. To understand the neural processing of the visual stimulus, a variety of visual stimuli, both natural and artificial images/videos are presented to the animal and respective neural responses are recorded from primary, secondary and higher order visual regions. Furthermore, the complexity in these neural responses is theoretically captured by a variety of computational models [4].

Although these deep learning models have some biological significance [5, 7] yet their usage has been majorly dedicated towards non-biological applications in the past.

In the recent years, deep neural networks have emerged as a strong candidate to predict the neural responses in primate's visual cortex [5, 6] but their application in modeling rodent datasets is quite scarce [7], partly due to the shortage of large neural datasets. More recently, deep learning-based approaches are developed to explore the mouse brain datasets [8, 9]. In this regard, Allen Brain Institute has been making critical contributions to the field by generating large publicly available experimental datasets from the mouse visual cortex [10]. This includes data collected using the state-of-the-art Neuropixel probe [11, 12]. These are electrical probes packed with hundreds of channels, allowing them to record from different layers of cortex as well as sub-thalamic regions. Each channel of the probe measures the extracellular potentials from which neuronal activities can be extracted. These datasets include the functional traces of individual spikes at the single neuronal level, that are recorded during the presentation of a variety of artificial (drifted gratings, orientated bars, etc.) and natural (bear, trees, cheetah, etc.) visual stimuli to the mice for a repeated number of trials.

In this study, we exploit the Neuropixel dataset [12] from Allen Brain Institute that contains spike responses of hundreds of neurons from mouse visual cortex along with a few sub-cortical brain regions. Utilizing these neural responses, we develop a novel approach of generating the *neural image dataset* for DNN by weight averaging each pixel of a given image (input stimulus) with the spiking response of the neurons (explained in the methods' section). We use neural image dataset to train and test the performance of a DNN in order to predict the input stimulus class (natural vs. artificial). We predict the stimulus class by applying transfer learning on a trained GoogleNet model [13] thereby training the DNN on neural image dataset. The classification accuracy of a DNN in decoding natural and artificial stimuli is recorded high both within and across animals.

The classical study by Hubel and Wiesel [20] demonstrated that each neuron in the primate visual cortex has a preferred stimuli or receptive field, and this principle is found to be conserved across mammalian species. In our study, we have expanded this concept to generate a *neural image* based on the preferred stimuli, and surprisingly found that decoding the neural responses can be carried out by using these neural images, or 'sum of preferred stimuli' across neural populations in a rodent model. Due to the limitation of identifying and characterizing individual neurons and their connections, our study does not take into account the different response profiles in simple and complex cells or neural hierarchy. However, while the detailed mechanism of visual perception and classification in the mammalian visual cortex is yet to be uncovered, our study provides a potential approach to decode the neural responses across different individual organisms.

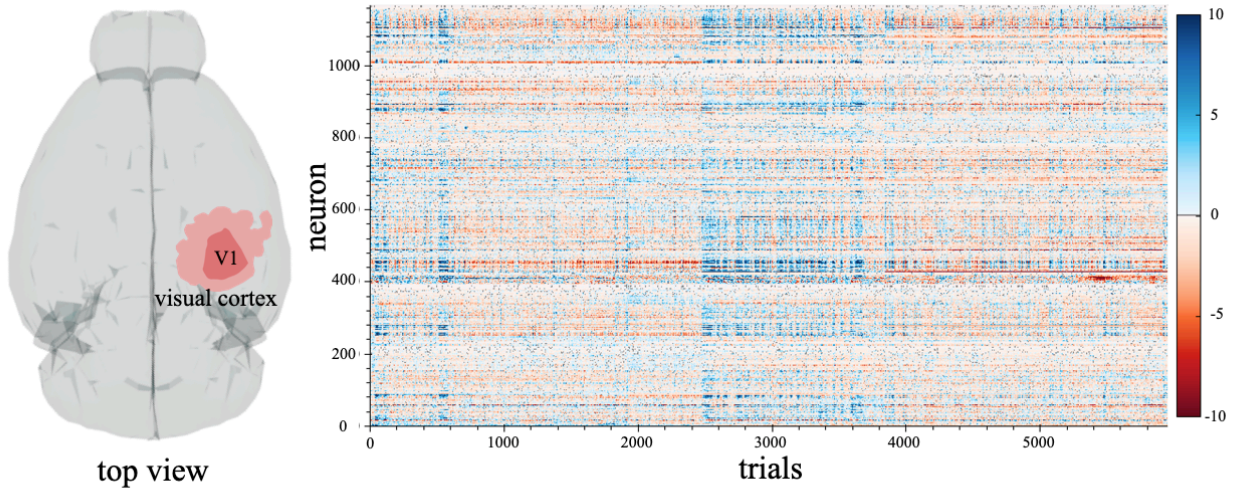

Fig. 1: Recording site and example data. Left panel, red area indicates the recording site. Right panel, a raster plot of neural responses to natural scenes for all neurons in one recording session. The activities are baseline-subtracted firing rates in Hz.

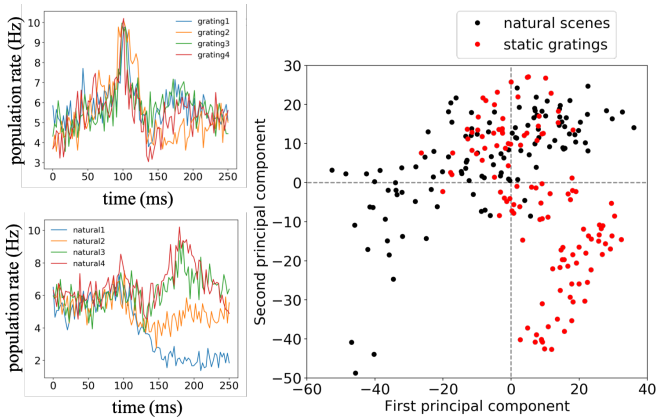

Fig. 2: Population spike patterns in response to different visual stimuli. Left panel (top) shows the neural responses of a sample neuron for four stimuli with different grating patterns. This neuron is sensitive to all the grating patterns. Bottom left shows the response of a sample neuron for four natural images stimuli, this neuron in area V1 provides a varying response for the variety of stimuli. Right panel shows the Principal Component Analysis (PCA) plot of neural responses for both static gratings and natural scenes stimuli. Each dot represents a population response profile to a particular visual stimulus. Responses are clustered by stimulus class with some major overlap.

This study opens a new dimension of discussion in utilizing the DNN models as a tool to understand the decoding and encoding principles in the mouse visual cortex.

## II. METHODS

In this section, development of neural image dataset from the Neuropixel dataset [12] is described along with a step by step explanation of training and testing paradigms for DNN. Figure 1 shows an overview of the Neuropixel dataset for natural scenes stimuli. Briefly, spiking data of a large population of neurons in primary visual cortex was obtained with Neuropixel probes. The animals were head-fixed throughout recording sessions and were allowed to run freely on a rotating disk. Different visual stimuli were presented to the animals across different sessions in a pseudo-random order. Two types of stimulus are of particular interest in this study: static gratings and natural scenes. The static gratings are composed of a set of sinusoidal gratings that vary in orientation, spatial frequency and phase. In a total of 6 orientations, 5 spatial frequencies, and 4 phases are used in the experiment whereas the natural scenes contain 118 images of a variety of living and non-living objects. Each variation of static gratings or natural scenes are presented to the animals for 250 ms, in blocks of approximately 50 trials. The data was acquired with OpenEphys GUI [14] at 30 kHz, and spikes were extracted using KiloSort [15]. All neural activities are subtracted from the baseline, which is the mean firing rate within 200 ms before the stimulus onset and subsequently normalized by its standard deviation. Figure 2 shows a couple of examples of averaged neural responses for multiple stimuli in both natural scenes and static gratings.

### A. Overview of neural responses

A simple approach to decode the visual stimulus from neural activities is to train the DNN with the raw neural responses. However, this approach does not take into account the information each neuron might be encoding. For instance, it can be speculated that a subset of neurons might prefer a slim bright "feature" at certain location of the visual field, and sustain a heightened response when the presented stimulus contains such a feature. In theory, the DNN should be able to learn such patterns of neural responses and probably represent them as hidden feature layers. In practice, however, the complexities of temporal dynamics of neural responses might make it a relatively hard problem for the neural network to acquire this kind of information. Thus, it is very intriguing to observe whether aggregating the neural responses over trials before the training of DNN could help improve the performance of prediction. The answer to this question will also reflect on whether the mouse visual cortex employs a fairly stable coding of features across time.

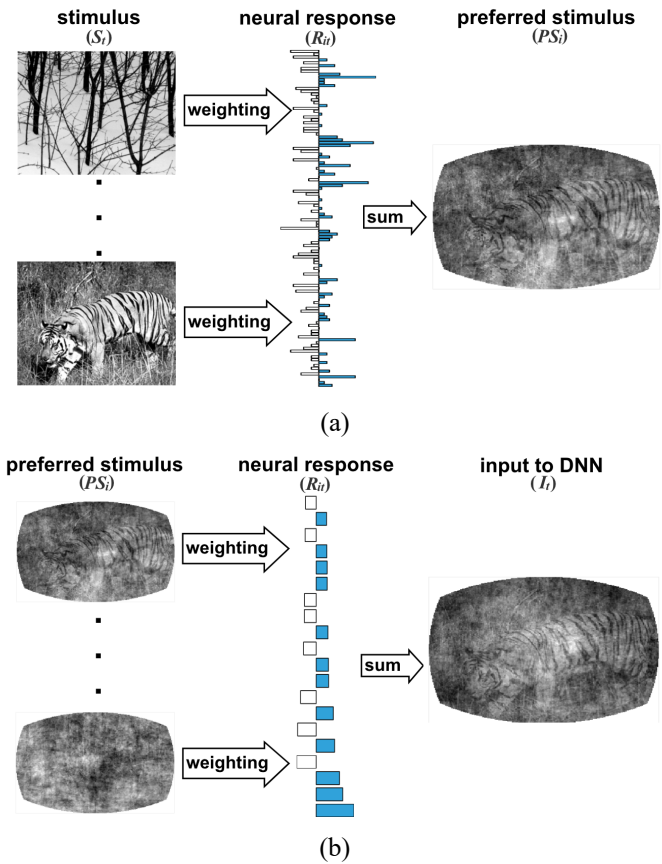

Fig. 3: Generation of the preferred stimulus and DNN inputs. (a) For a given neuron, we calculate the neural response to specific visual stimulus (natural images or static gratings) by taking the mean baseline-subtracted firing rate during the epoch of stimulus presentation. Then, the visual stimulus are weighted by the corresponding neural responses. The preferred stimulus of said neuron is then the sum of all the weighted stimulus. (b) For a given stimulus, we calculate the response from all recorded neurons by taking the mean baseline-subtracted firing rate during the epoch of stimulus presentation. Then, we weight the preferred stimulus of each neuron by their response to the said stimulus. We then use the weighted sum of all preferred stimulus as the input to DNN of the said stimulus.

Furthermore, aggregating responses for each neuron over trials enable us to carry out predictions across animals. Normally, there is no relationship between the neurons from one animal and those from the other, thus the weight matrix from a DNN trained with neural response of one animal usually cannot be generalized and used to predict stimulus for the other animal. However, if we generate a "preferred stimulus" for each neuron by aggregating their response over trials, those "preferred stimulus" presumably represent the functional role of each neuron, and can serve as a common basis across different animals. Thus, we ask the question whether DNN trained with the "preferred stimulus" from one animal can predict the stimulus for the other animal. The answer to this question will show whether the feature-based rules are common and relatively similar across different animals.

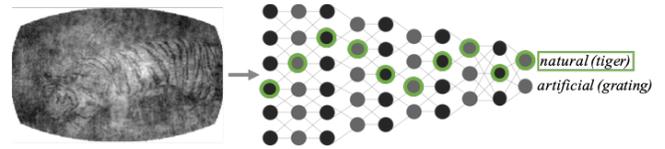

Fig. 4: Preferred stimulus (output of Figure 3 (b)) is fed as an input to the GoogleNet here shown as a DNN architecture with different patterns of activations to predict the stimulus (natural scenes – tiger) class.

### B. Measuring the preferred stimulus

To generate a preferred stimulus for each neuron, we superimpose different stimulus (artificial gratings or natural scenes), weighted by the neural response, across trials, using the following equation:

$$PS_i = \sum_{t=0}^{T} R_{it} \times S_t$$

where $i$ is the neuron index, $t$ is trial index, $PS_i$ is the preferred stimulus for neuron $i$, $R_{it}$ is the response of neuron $i$ at trial $t$ averaged across the duration of the trial, and $S_t$ is the stimulus presented at trial $t$. The procedure is visualized in Figure 3 (a).

### C. Training DNN with neural images

To train the DNN, we first develop novel images that represent how a population of neurons in the primary visual cortex encode each visual stimulus (herein 'neural image'). Then, the DNN is trained to perform classification task on those neural images. To construct a neural image that captures the response of a population of neurons to a stimulus presentation, we superimpose the preferred stimulus of each neuron weighted by neural response across neurons (Figure 3 (b)), using the following equation:

$$I_t = \sum_{i=0}^{N} R_{it} \times PS_i$$

where $I_t$ is the input to the DNN on trial $t$, $R_{it}$ is the response of neuron $i$ at trial $t$ averaged across the duration of the trial, and $PS_i$ is the preferred stimulus of neuron $i$. We then use part of $I_t$s as training set to the DNN (Figure 4) and test the prediction on another set of $I_t$s, for the classification task. Note that the equation does not imply animal identity at all. We can thus use the set of $I_t$s from one animal to train the DNN, and test whether the trained network can make good predictions with inputs ($I_t$s) from another animal.

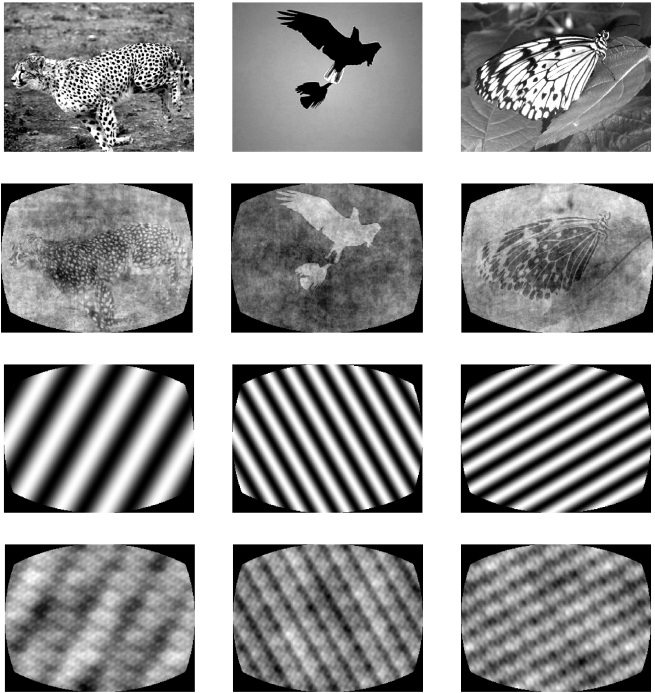

Fig. 5: Sample of natural and artificial image samples from input stimuli and neural images dataset. Rows #1 and #3 show a sample of stimulus images from natural and artificial object categories respectively. Rows #2 and #4 show the corresponding neural images from each of the presented stimulus. It can be observed that the neurons in visual cortex are able to encode the distinct features in natural scenes and static gratings.

*D. Transfer learning on DNN*

After generating the neural dataset, we divided the neural response images into two categories: *natural* (real world stimulus images) and *artificial* (static gratings with different orientation and frequency). Figure 5 shows a few examples of input stimulus and the corresponding neural images. We randomly choose 2/3rd of the dataset for training and 1/3rd for testing after cross-validation. We optimize a high performance deep neural network (Inception/GoogleNet) to train and test the neural images for predicting the stimulus class. GoogleNet is pre-trained on the ImageNet [16] dataset and we apply transfer learning [17] on the trained model thereby retraining the network on *natural* and *artificial* classes of neural image dataset.

## III. RESULTS

After generating the neural image dataset, we conducted two types of experiments to perform the prediction of stimulus class: i) training and testing the DNN on the neural images within a single animal and ii) train the DNN on the neural images of a set of animals and testing on another set. Results of each experiment are presented in detail below.

*A. Predicting stimuli within animals*

In Neuropixel data from Allen Brain Institute, neural responses were recorded from 9 adult animals namely S0-S4 and M0-M3. We utilize this neural data from visual cortex of all of these animals and decode the neural responses within each animal by training and testing the DNN on neural images for natural and artificial stimuli within the same animals. Prior to testing the performance of DNN, we trained a simple GLM (generalized linear model) on the neural dataset to check the baseline and the average classification score of 83% was achieved. On average, DNN is able to decode the neural responses with as much as 96% confidence score. In some animals, e.g. S0, classification accuracy has dropped to 89% whereas it is perfect in M0 animal as shown in Figure 6 (a). This gives an intuition that the neural responses within the animals are quite consistent across different object categories and the visual neurons are able to spatially encode the distinct features of natural and artificial images in a precise manner.

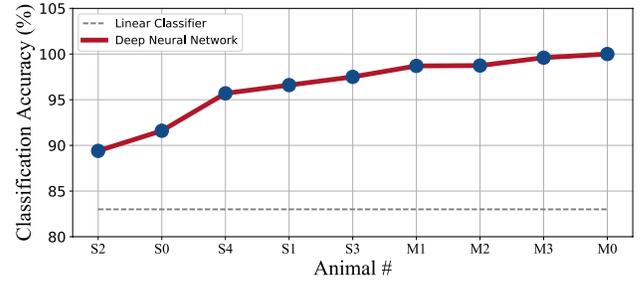

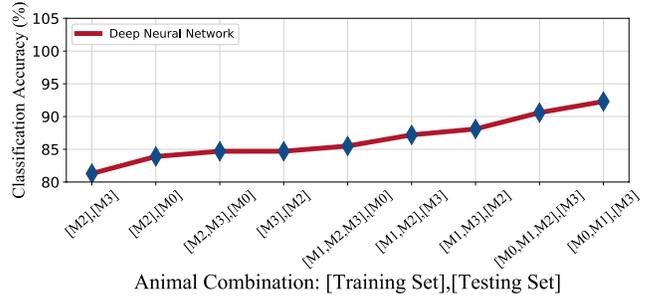

Fig. 6: Prediction of stimulus class by training and testing the DNN on neural image datasets. (a) shows the classification performance on training and testing the DNN within a single animal where y-axis labels represent the animal numbers. DNN is able to predict the stimulus class with a very high classification score. (b) DNN is trained and tested on a diverse combinations of animals, it can be observed that neural responses for a stimulus class are consistent across animals with a DNN classification score as high as 91.2 % by training on M0 and M1 animals and testing on M3.

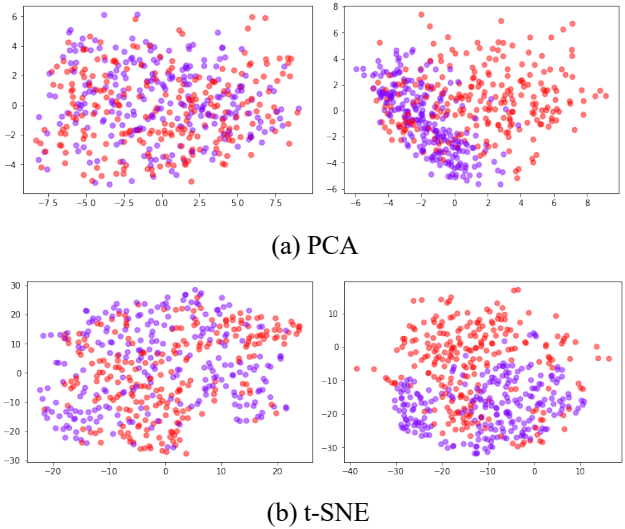

(a) PCA

(b) t-SNE

Fig. 7: Dimensionality reduction to visualize the diversity of neural image datasets of different animals for training and testing on DNN. Horizontal axis shows the first principal component and vertical axis shows the second principal component. (a) PCA is applied on two different animal combinations of training and testing sets. (b) shows the same datasets visualized by t-SNE. Right dataset seems to show sparse clustering of natural scenes (red) and static gratings (purple) as compared to the left one. These two datasets also provide the maximum (91.2%) on right and minimum (82%) on left classification scores in multiple animal plots in Figure 6 (b).

### B. Dimensionality reduction on neural image dataset

To check the performance of DNN across multiple animals, we selected 4 out of 9 recorded animals and generated a variety of combinations of neural image dataset for both training and testing e.g. [M0,M1],[M3] — training the DNN on neural images of M0 and M1 and testing on M3 animal, etc. Before training the network on DNN, we visualized and observed the variability in the neural image datasets by applying two types of dimensionality reduction techniques on the neural image datasets: Principal Component Analysis (PCA) [18] and T-distributed Stochastic Neighbor Embedding (t-SNE) [19]. Figure 7 (left) shows the PCA and t-SNE plots on [M2],[M3] dataset and Figure 7 (right) shows the PCA and t-SNE plots on [M0,M1],[M3] dataset. Purple points show the neural images from the natural (static scenes) stimuli whereas the red points represent the neural images of artificial (static gratings) stimuli. There seems to be an emergence of two clusters (natural and artificial stimuli) overall as shown by PCA and t-SNE results on the [M0,M1],[M3] animal set as compared to [M2],[M3]. This finding indicates that neural responses are encoding the stimulus in a differential manner.

### C. Predicting stimuli across multiple animals

After testing the performance of individual animals, we wanted to check the consistency in neural responses across different animals. To test this idea, we picked the top 4 animals (M0, M1, M2, M3) from the single animal category and made a random combination of training and testing samples from different animals as shown in Figure 6 (b). Classification scores are presented in an ascending order. Training the network on M2 animal and testing on M3 gave the minimum score. Similarly PCA and t-SNE plots in Figure 7 (left) show a huge overlap in the sample space for the same dataset. Another combination [M0,M1],[M3] gave the highest classification score which also provides a rather sparse clustering in PCA and t-SNE plots as shown in Figure 7 (right). Remarkably, the neural space of animal M0 and M1 is able to capture the complexity of neural responses in M3 with a very high classification score (91.2 %).

## IV. DISCUSSION

Our findings suggest that neural space in area V1 of mouse is able to capture the complexity of feature space of unique object structures. These results are consistent within and across different animals. It is observed that a powerful DNN is able to decode the neural responses in mouse visual cortex for natural scenes and static gratings class. We noticed that the neural responses for the two classes of stimuli are distinguishable within animal, which could be attributed to a simple alternative hypothesis such as the different level of activities may be coming from different sub-regions of visual cortex. However, an increase in classification performance when adopting neural images as training sets, as well as persistent high performance of classification across animals, indicates that our approach adds in more information that cannot be captured by simple differential population activities within animals. Our study suggests that such information might be closely related to the tuning property of each neuron regarding visual stimulus. Although the neural responses are distinct for stimulus class but it also remains a bit unclear whether and how much is this driving the performance of DNN in classifying the neural images. One possibility remains that the information about the original stimuli themselves, rather than the neural activities, drives the classifications.

We investigated it further by conducting an independent analysis not based on the weighted superimposition of the original stimuli (neural images), but exclusively based on temporal neural responses fed onto GoogleNet (data not shown). In this rather conventional way, we took neural spikes of each neuron over a time bin, averaged over 50 stim presentations and baseline subtracted and normalized by standard deviation. Then we generated 2-dimensional image of a time series (x-axis) of individual neurons (y-axis) whose spike rate is represented as pixels in gray scale, for each visual stimulus that belongs to either natural scenes or static gratings. These images were used for training and testing the DNN in the same way as the superimposed neural images were used for our experiments. Strikingly, training the same dataset pairs for training and testing generated slightly lower but still considerably high classification score (CA). In addition, there was a high linear correlation of CA between the two analyses for the same mouse dataset pairs, i.e. the [M0, M1],[M3] pair scored about 91% (highest) in the superimposed image analysis and 72% (highest) in the neural spike analysis. This result excludes the possibility that the high classification performance of neural images is purely coming from the power of DNN to distinguish the original stimuli class.

Visual stimuli are complex composite of multiple parameters of individual elements, including orientation, color, brightness, motion, etc. Natural scene stimuli that are used in this experiment may contain some individual elements of static gratings that are embedded within the images. It is likely that within the V1 area, the same subset of neurons responded both to natural scenes and static gratings, depending on the elements of the visual stimuli. It is further possible that since natural scenes are generally more complex so their response profiles can be influenced by additional factors such as excitation or inhibition by other wired neurons as well as neural modulation (i.e. arousal by a predator stimulus). Since, we are investigating a new decoding approach from a population dynamics of mouse primary visual cortex, we chose the most distinct types of visual stimuli: natural scenes and static gratings. Our findings show a promising result which can be expanded to more specific visual stimuli with controlled complexity, i.e. static gratings of 0° vs 90° orientation.

Furthermore, we also tested the performance of DNN in decoding the sub-classes in natural scenes dataset. DNN was able to decode the neural patterns with as much as 70% classification score for predator vs non-predator class (data not shown). However, the dataset is relatively small in this regard and DNN needs a lot of training samples. In any case, if DNN is able to separate the neural response patterns in its feature space then it is also indicative of the fact that such networks can also be used as biologically-inspired models to understand the architecture and functionality of mouse visual cortex.

## V. Conclusion

In this study, we utilize the power of a deep neural network to demonstrate that the neurons in mouse visual cortex are able to encode the natural (scenes) and artificial (static gratings) stimuli in a distinct manner. We tested this idea on Neuropixel dataset which consists of responses of hundreds of neurons in mice visual cortex. Our findings propose a novel methodology to generate the neural image dataset from the neural responses of mouse visual cortex. The approach of superimposing the input stimulus by the weight average of the neural responses across trials (for their preferred stimulus) offers a powerful method to analyze and visualize the global activity of all the neurons against a single stimulus. Furthermore, by dint of our study, we are also able to demonstrate that neural responses in mouse visual cortex are consistent within and across different animals for a large set of natural scenes and static gratings stimuli. We evidenced this by training the DNN on neural images/responses of natural and artificial stimuli collected from one group of animals and tested on other groups.

## VI. Future Work

Neuropixel data also contains recordings from sub-cortical regions such as thalamus and hippocampus. It will be interesting to explore if the feature processing or object classification of visual stimulus is already computed by the thalamic neurons before the information is propagated to area V1. Similarly, hippocampal neurons may encode the visual stimuli in the memory circuit, and provide distinct responses before the onset of stimulus. Furthermore, we aim to explore if the neurons in V1 area of mice encode visual information for sub-classes in natural vs artificial categories in a unique manner e.g. birds, reptiles, horizontal gratings, vertical gratings, etc. Future study also aims to compare the data according to the arousal state of the animal, which can profoundly change the neural response to the same visual stimuli. All of these theories can be tested by training the DNN on neural images collected from different brain regions, object categories, and the state of the animal.

## Acknowledgment

We would like to acknowledge the financial support from the Simons Foundation Collaboration for the Global Brain. We would also like to thank R. Aoki (RIKEN Center for Brain Science) for the insightful discussions and for providing critical feedback during this study. We would also like to acknowledge all the technical support of staff, faculty and organizers of Summer Workshop on the Dynamic Brain (SWDB) 2018. SWDB was co-organized by the Allen Brain Institute in Seattle, WA and Computational Neuroscience program of the University of Washington, Seattle, WA. In particular, we would like to extend our gratitude to M. Buice, S. Olsen and C. Koch (Allen Institute for Brain Sciences) and A. Fairhall and E.S. Brown (University of Washington, Seattle) for providing a constructive feedback about this work. Finally, we would like to thank all the participants of SWDB 2018 workshop for their valuable suggestions and feedback.